\documentclass[aip, cha, reprint, twocolumn, groupedaddress, floatfix]{revtex4-1}

\usepackage{dcolumn}
\usepackage{amsmath, amssymb}
\usepackage[english]{babel}
\usepackage[latin1]{inputenc}
\usepackage{xy}
\usepackage{graphicx}
\usepackage{verbatim}
\usepackage{enumerate}
\usepackage{url}
\usepackage{bm}
\usepackage{times}

\usepackage{xcolor}

\newcommand{\fgdir}{}

\newcommand{\aln}{\alpha_n}
\newcommand{\als}{\alpha_s}

\newcommand{\SSz}{\mathbf{SS}_0}
\newcommand{\SSp}{\mathbf{SS}_\pi}
\newcommand{\Ip}{\textrm{I}}

\DeclareMathOperator{\Rep}{Re}
\DeclareMathOperator{\Imp}{Im}
\renewcommand{\Re}{\Rep}
\renewcommand{\Im}{\Imp}

\newcommand{\OP}{Z}
\newcommand{\BP}{z}

\newcommand{\psiCM}[2]{\psi_{#2}^{(#1)}}

\newcommand{\twocol}[1]{#1}
\newcommand{\half}{}

\newcommand{\maxdim}{N}

\newcommand{\ud}{\mathrm{d}}
\newcommand{\udi}{\,\ud}

\newcommand{\R}{\mathbb{R}}

\newcommand{\Z}{\mathbb{Z}}

\newcommand{\C}{\mathbb{C}}

\newcommand{\lmax}{\lambda_{\text{max}}}

\newcommand{\Tor}{\mathbf{T}}

\newcommand{\Tornn}{\Tor^{2\maxdim}}

\newcommand{\abs}[1]{\left|#1\right|}

\newcommand{\sset}[1]{\left\lbrace #1\right\rbrace}

\begin{document}

\preprint{AIP/123-QED}


\title{Chaos in Kuramoto oscillator networks}%
\author{Christian Bick${}^{a,b}$}%
\author{Mark J.~Panaggio${}^{c}$}%
\author{Erik A.~Martens${}^{d,e,f}$}%

\affiliation{%
\twocol{\mbox}
{${}^a$Department of Mathematics and Centre for Systems Dynamics and Control, University of Exeter, Exeter EX4~4QF, UK}\\
\twocol{\mbox}
{${}^b$Oxford Centre for Industrial and Applied Mathematics, Mathematical Institute, University of Oxford, Oxford OX2~6GG, UK}\\
\twocol{\mbox}
{${}^c$Department of Mathematics, Hillsdale College, 33 E College Street, Hillsdale, MI 49242, USA}\\
\twocol{\mbox}
{${}^d$Department of Applied Mathematics and Computer Science, Technical University of Denmark, 2800 Kgs.~Lyngby, Denmark}\\
\twocol{\mbox}
{${}^e$Department of Biomedical Sciences, University of Copenhagen, Blegdamsvej 3, 2200 Copenhagen, Denmark}\\
\twocol{\mbox}
{${}^f$Department of Mathematical Sciences, University of Copenhagen, Universitetsparken 5, 2200 Copenhagen, Denmark}
}
\date{\today}


\begin{abstract}%
Kuramoto oscillators are widely used to explain collective phenomena in networks of coupled oscillatory units. We show that simple networks of two populations with a generic coupling scheme, where both coupling strengths and phase lags between and within populations are distinct, can exhibit chaotic dynamics as conjectured by Ott and Antonsen [Chaos, 18, 037113 (2008)]. These chaotic mean-field dynamics arise universally across network size, from the continuum limit of infinitely many oscillators down to very small networks with just two oscillators per population. Hence, complicated dynamics are expected even in the simplest description of oscillator networks. 
\end{abstract}

\maketitle

\begin{quotation}
Phase oscillator models---such as Kuramoto's model---have been instrumental to understand synchronization phenomena in networks of identical (or almost identical) coupled oscillators. What coupling properties are necessary such that these model systems can exhibit chaotic dynamics? While heterogeneity can induce microscopic chaotic fluctuations for globally and identically coupled phase oscillators, the chaos vanishes as the number of oscillators goes to infinity. Here we show that simple networks of two identical populations of identical (and almost identical) phase oscillators with sinusoidal interactions of Kuramoto type support chaos. These chaotic dynamics appear in both the smallest possible networks of just four oscillators and in macroscopic descriptions of infinitely large networks for similar parameter values; for these parameter values the network has attractive as well as repulsive interactions.
Hence, neither oscillator heterogeneity, amplitude variations, nor more complicated interactions are necessary to see chaos in coupled oscillator networks.
\end{quotation}

\section{Introduction}
The Kuramoto phase model~\cite{Kuramoto} and its generalization by Sakaguchi~\cite{Sakaguchi1986} are widely used to understand synchronization and other collective phenomena in weakly coupled oscillator networks in physics and biology~\cite{Acebron2005, Rodrigues2016}. 
Networks of globally coupled identical Kuramoto oscillators cannot exhibit chaotic dynamics because degeneracy leads to dynamics that are effectively two-dimensional~\cite{Watanabe1993}. Moreover, chaos in finite networks of globally coupled nonidentical units vanishes in the continuum limit of infinitely many oscillators~\cite{Popovych2005}.
Hence, a decade ago, Ott and Antonsen conjectured in their seminal paper~\cite{Ott2008} that networks of two or more populations---where interactions are all-to-all but distinct between and within populations---could exhibit chaotic mean-field dynamics, both in the continuum limit and in finite networks. 
However, the dynamics which have been observed for coupled populations of Kuramoto oscillators yield periodic and quasiperiodic motions of the mean field in the continuum limit~\cite{Abrams2008, Pikovsky2011}.

In this paper, we report macroscopic mean-field chaos for two populations of~$\maxdim$ Kuramoto phase oscillators and their continuum limit $\maxdim\to\infty$. More specifically, we consider oscillator networks where the phase ${\theta}_{\sigma,k}\in\Tor:=\R/2\pi\Z$ of oscillator~$k\in\sset{1,\dotsc,\maxdim}$ in population $\sigma\in\sset{1,2}$ evolves according to
\begin{align}\label{eq:PhaseOsc}
 \dot{\theta}_{\sigma,k} &= \omega_{\sigma,k}+\sum_{\tau=1}^2 \frac{K_{\sigma\tau}}{\maxdim}\sum_{j=1}^{\maxdim}\sin\!\left(\theta_{\tau,j}-\theta_{\sigma, k}-\alpha_{\sigma\tau}\right);
\end{align}
the intrinsic frequencies~$\omega_{\sigma,k}$ are sampled from a Lorentzian distribution with half-width-at-half-maximum~$\Delta$~\footnote{Due to the rotational invariance of the Kuramoto equations~\eqref{eq:PhaseOsc} we have assumed the Lorentzian to be centered at zero without loss of generality.} and~$K_{\sigma\tau}$ and $\alpha_{\sigma\tau}$ are the coupling strength and phase lag between populations~$\sigma$ and~$\tau$. While~\eqref{eq:PhaseOsc} has been extensively studied for networks with identical phase lags $\alpha_{\sigma\tau}=\alpha$~\cite{Sakaguchi1986, Ott2008, Abrams2008, Panaggio2015}, we find here that chaotic dynamics arise in the generic situation where both coupling strength~$K_{\sigma\tau}$ and phase lags~$\alpha_{\sigma\tau}$ are distinct~\cite{Martens2016, Choe2016}. Remarkably, the chaotic dynamics not only appear in the continuum limit $\maxdim\to\infty$ of~\eqref{eq:PhaseOsc} and for large~$\maxdim$, but also in small networks down to just $\maxdim=2$ oscillators per population. 
First, our results provide a positive answer to Ott and Antonsen's conjectures for minimal networks of two populations.
Second, neither oscillator heterogeneity, amplitude variations, the influence of fast oscillations, nonautonomous forcing, nor higher-order interactions or derivatives (see for example Refs.~\onlinecite{Marvel2009, Bick2011, So2011, Komarov2013a, Pazo2014, Bick2016b, Olmi2015, Maistrenko2016}) are necessary to observe chaos. Hence, we anticipate that such chaotic phase dynamics arise in a large number of real-world systems~\cite{Martens2013, Bick2017}.

\section{Chaotic Mean-Field Dynamics in the Continuum Limit.}%
Each oscillator of the network~\eqref{eq:PhaseOsc} is driven by a common mean field which depends on the Kuramoto order parameter
\begin{equation}\label{eq:OPdisc}
\OP_\sigma = r_\sigma e^{i\phi_\sigma}  = \frac{1}{\maxdim}\sum_{j=1}^\maxdim e^{i\theta_{\sigma,j}}
\end{equation}
of population~$\sigma$; here $i = \sqrt{-1}$. The order parameter encodes the level of synchrony of the population: $\abs{{\OP}_{\sigma}}=r_\sigma=1$ if and only if population~$\sigma$ is fully phase synchronized. Write $\als:=\alpha_{\sigma\sigma}$, $k_s:=K_{\sigma\sigma}$ for the self-coupling strength and phase lag, and $k_n:=K_{12}=K_{21}$, $\aln:=\alpha_{12}=\alpha_{21}$ for the neighbor-coupling strength and phase lag. 
By rescaling time appropriately we set $k_s + k_n = 1$ and parametrize the deviation $A = k_s - k_n$ of coupling strengths.
This yields the complex coupling parameters $c_s = c_s(\als, A):=k_s e^{-i\alpha_s}$, $c_n=c_n(\aln, A):=k_n e^{-i\alpha_n}$. Now
\begin{equation}\label{eq:Field}
H_\sigma = c_s\OP_\sigma + c_n\OP_{\tau},
\end{equation}
where $\tau=2$ if $\sigma=1$ and $\tau=1$ if $\sigma=2$, drives the evolution of population $\sigma$ since~\eqref{eq:PhaseOsc} can be rewritten as
\begin{align}\label{eq:PhaseOscField}
\dot{\theta}_{\sigma,k} = \omega_{\sigma,k}+\Im(H_\sigma e^{-i\theta_{\sigma,k}}).
\end{align}

In the continuum limit, the system~\eqref{eq:PhaseOscField} is described by the evolution of the probability density~$f_{\sigma}(\theta, t)$ for an oscillator of population~$\sigma$ to be at $\theta\in\Tor$ at time~$t$. In the limit, the order parameter~\eqref{eq:OPdisc} of population~$\sigma$ is ${\OP}_{\sigma}(t) = r_\sigma(t) e^{i\phi_\sigma(t)}  =\int_{0}^{2\pi}e^{i\theta}f_{\sigma}(\theta, t)\udi\theta$. For $w\in\C$ let~$\bar w$ denote its complex conjugate. Ott and Antonsen~\cite{Ott2008} showed that there is an invariant manifold of densities~$f_{\sigma}$ on which the dynamics are determined by
\begin{align}\label{eq:ComplxEqns}
\dot \OP_\sigma &=  -\Delta \OP_\sigma+\frac{1}{2}H_\sigma - \frac{1}{2}\bar H_\sigma{\OP}_\sigma^2.
\end{align}
Since these equations are symmetric by shifting phases by a constant angle, we introduce the phase difference $\psi=\phi_2-\phi_1$ to obtain the three-dimensional system%
{\allowdisplaybreaks
\begin{subequations}\label{eq:cylgov}
\begin{align}
 \label{eq:cylgova}
 \dot{r}_1&= -\Delta r_1+\frac{1-r_1^2}{2}\left(r_1\Re(c_s)+r_2\Re(\bar c_n e^{-i\psi})\right)\\
 \label{eq:cylgovb}
 \dot{r}_2&= -\Delta r_2+\frac{1-r_2^2}{2}\left(r_2\Re(c_s)+r_1\Re(\bar c_n e^{i\psi})\right)\\
 \begin{split}
 \label{eq:cylgovc}
 \dot{\psi}&=
 \frac{1+r_1^2}{2r_1}\left(r_1\Im(\bar c_s)+r_2\Im(\bar c_n e^{-i\psi})\right)
 \twocol{\\&\qquad}
 -\frac{1+r_2^2}{2r_2}\left(r_2\Im(\bar c_s)+r_1\Im(\bar c_ne^{i\psi})\right)
 \end{split}
\end{align}
\end{subequations}}%
which describes the dynamics of two populations through their level of synchrony $0<r_1, r_2\leq 1$ and $\psi\in[0, 2\pi)$. The equilibrium $\SSz=(1, 1, 0)$ corresponds to full (phase) synchrony, $\SSp=(1, 1, \pi)$ to a two cluster solution where the clusters are in anti-phase, and $\Ip = (0, 0, *)$ denotes completely incoherent configurations with $\OP_1=\OP_2=0$. Moreover, there is a time-reversal symmetry for $(\als, \aln) = (\frac{\pi}{2}, 0)$; cf.~Ref.~\onlinecite{Martens2016} for details.

\begin{figure}
\includegraphics[width=\half\linewidth]{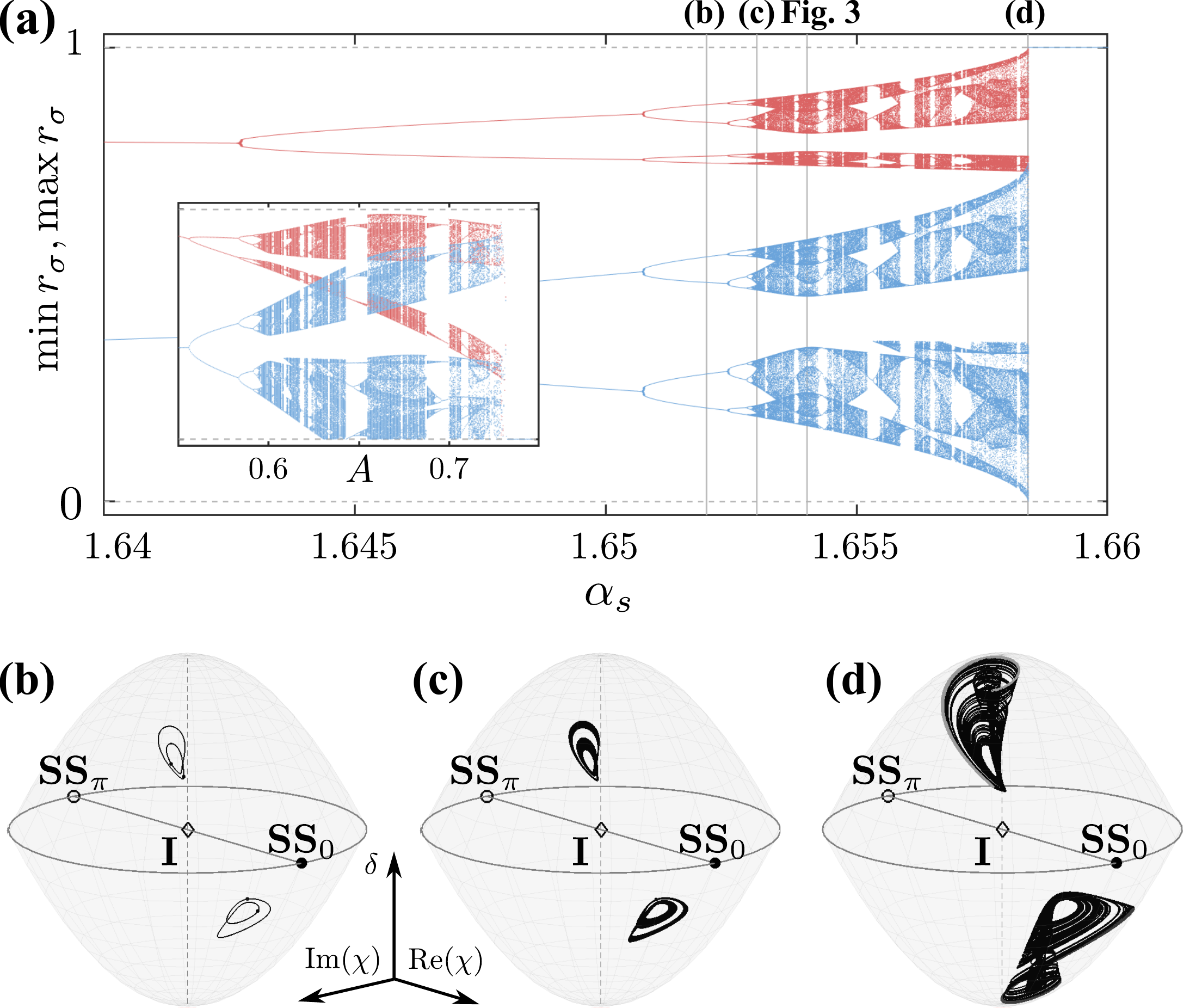}
\caption{\label{fig:Chaos}Chaotic attractors arise for the mean-field dynamics~\eqref{eq:cylgov} for $A=0.7$ and fixed $\aln=0.44$. 
Panel~(a) shows the local maxima and minima of~$r_\sigma=\abs{\OP_\sigma}$ (red/blue). A stable equilibrium where both populations partly synchronized, $0<r_1, r_2<1$, looses stability in a Hopf bifurcation as~$\alpha_s$ is increased. The emerging periodic orbit goes through a period doubling cascade to chaos. The chaotic attractor is eventually destroyed as it approaches the invariant surface $r_\sigma=1$ and $r_\sigma=0$ (dashed lines). In the inset,~$A$ is varied while $\als = 1.654$ is fixed. Initial conditions were continued quasi-adiabatically as parameters are varied.
Panels~(b--d) show two symmetry related trajectories (black curves) for the parameter values highlighted in Panel~(a) by vertical lines: (b)~after the first period doubling, $\als=1.652$, (c)~after the first transition to chaos, $\als=1.653$, and (d)~just before the crisis, $\als=1.6584$. 
In the projection $(\chi, \delta)= (\OP_1\bar\OP_2, \abs{\OP_1}^2-\abs{\OP_2}^2)$, the permutational symmetry of the populations corresponds to the map $(\chi, \delta)\mapsto(\bar\chi, -\delta)$. Consequently, the invariant surfaces $r_1=1$ (shaded, top) and $r_2=1$ (bottom) intersect in the unit circle on the $\chi$-plane (circular line). Points on the attractor in close proximity to these invariant surfaces are highlighted in gray.
}
\end{figure}

Chaotic attractors arise in the mean-field dynamics~\eqref{eq:cylgov} of the continuum limit. First, consider identical oscillators, $\Delta=0$. The bifurcation diagram in Fig.~\ref{fig:Chaos}(a) shows that chaos arises through a period-doubling cascade of periodic orbits; here we fixed $A=0.7$ but there is a range of~$A$ for which there are chaotic dynamics (see inset of Fig.~\ref{fig:Chaos}(a)). The periodic orbits bifurcate from a stable equilibrium with $0<r_1,r_2<1$ which gains stability in a transcritical bifurcation where a ``classical chimera'' with $r_\sigma<r_\tau=1$ becomes unstable~\cite{Martens2016,Choe2016}.
As~$\alpha_s$ is increased, the chaotic attractors are destroyed as they approach the invariant surfaces $r_\sigma=1$ where one of the populations is phase-synchronized. The system symmetry $(r_1, r_2, \psi)\mapsto(r_2, r_1, -\psi)$ implies the existence of two attractors which are related by symmetry. Hence, there is multistability of the fully synchronized equilibrium~$\SSz$ and two chaotic attractors. Note that the phase difference of the mean fields~$\psi$ is bounded~(see Fig.~\ref{fig:Chaos}(b--d)), that is, the centroids of the order parameters~$\OP_\sigma$ do not rotate relative to one another. 

\begin{figure}
\includegraphics[width=\half\linewidth]{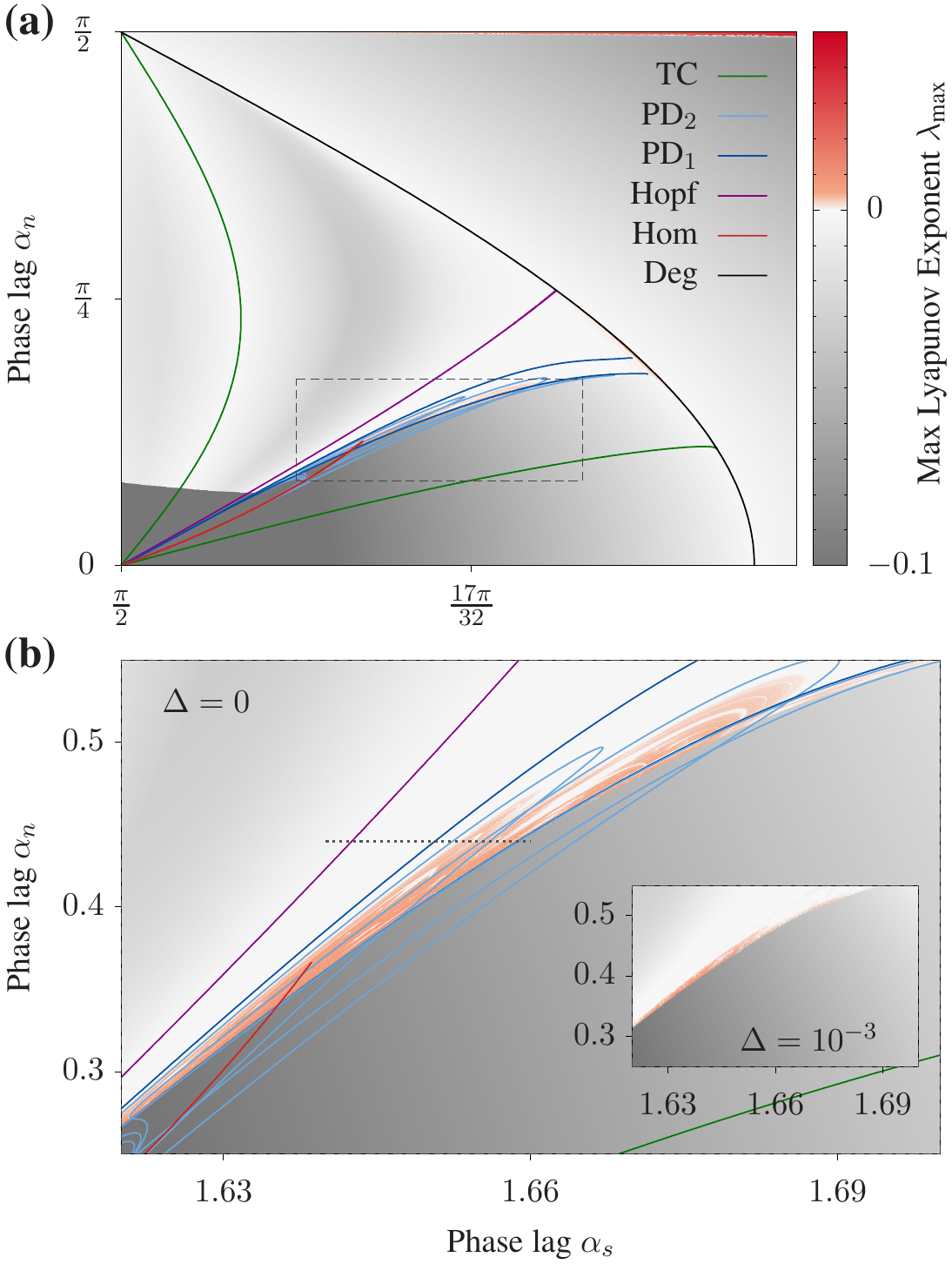}
\caption{\label{fig:Continuation}The mean-field equations~\eqref{eq:cylgov} show positive Lyapunov exponents (coloring) in a region of $(\als, \aln)$-parameter space for $A=0.7$. The system was integrated numerically from the fixed initial condition~$(r_1(0), r_2(0), \psi(0))=(0.8601, 0.4581, 1.1815)$. 
Panel~(a) shows the maximal Lyapunov exponents overlaid with two-parameter bifurcation lines: the transcritical (TC), Hopf, and first period-doubling (PD$_1$) lines emanate from $(\als, \aln) = (\frac{\pi}{2}, 0)$ and end in the degenerate bifurcation (Deg) where~$\SSz$ and~$\SSp$ swap stability~\cite{Martens2016}. 
Panel~(b) shows a magnification of the region highlighted in Panel~(a) where positive Lyapunov exponents arise (red color); a dotted line indicates the parameter range shown in Fig.~\ref{fig:Chaos}. Chaotic regions are bounded by ``lobes'' of second period-doubling~PD$_2$ lines.
The inset shows that positive Lyapunov exponents persist in the same parameter range for nonidentical oscillators with a nontrivial distribution of intrinsic frequencies $\Delta>0$.
}
\end{figure}

To quantify the chaotic dynamics we calculate the maximal Lyapunov exponents~$\lmax$ for the mean-field equations~\eqref{eq:cylgov}. Fig.~\ref{fig:Continuation} shows a region in $(\als, \aln)$-parameter space where the maximal Lyapunov exponents are positive. Numerical continuation of the bifurcations shown in Fig.~\ref{fig:Chaos} in the parameter plane using AUTO~\cite{Doedel2000} shows that the chaotic region is organized into multiple ``lobes'' which are bounded by period-doubling curves (PD$_2$ in Fig.~\ref{fig:Continuation}). Moreover, multiple bifurcation lines---including period doubling and a homoclinic bifurcation---end in the point $(\als, \aln) = (\frac{\pi}{2}, 0)$ where the system has a time-reversal symmetry. Hence, these parameter values appear to organize the bifurcations. 

The chaotic dynamics in the continuum limit persist for nonidentical oscillators, $\Delta>0$, as shown in Fig.~\ref{fig:Continuation}(b). At the same time, the invariant manifold of probability densities described by Ott and Antonsen attracts a class of probability densities $f_{\sigma}(\theta, t)$ for $\Delta>0$~\cite{Ott2009, Mirollo2012}. Hence, the long-term dynamics of the continuum limit of~\eqref{eq:PhaseOsc} will exhibit chaotic mean-field dynamics for a range of initial oscillator distributions.

\section{Chaotic Dynamics in Finite Networks.}%
The networks dynamics~\eqref{eq:PhaseOsc} of two finite populations of $\maxdim>3$ identical oscillators, $\omega_{\sigma,k}=\omega$, can be described exactly in terms of collective variables~\cite{Watanabe1993, Marvel2009, Pikovsky2008}. (We assume $\omega=0$ without loss of generality.) Then the phase space~$\Tornn$ of~\eqref{eq:PhaseOsc} is foliated by six-dimensional leafs, each of which is determined by constants of motion~$\psiCM{\sigma}{k}$, $k=1, \dotsc, \maxdim$ ($\maxdim-3$ are independent). The dynamics of population~$\sigma=1,2$ on each leaf are given by the evolution of its bunch amplitude~$\rho_\sigma$, bunch phase~$\Phi_\sigma$, and phase distribution variable~$\Psi_\sigma$. Write $z_\sigma=\rho_\sigma e^{i\Phi_\sigma}$. The bunch variables relate to the order parameter~\eqref{eq:OPdisc} through 
$\OP_\sigma = \BP_\sigma\gamma_\sigma$
where
\[\gamma_\sigma = \frac{1}{\maxdim\rho_\sigma}\sum_{j=1}^\maxdim\frac{\rho_\sigma e^{i\Psi_\sigma} + e^{i\psiCM{\sigma}{j}}}{e^{i\Psi_\sigma} + \rho_\sigma e^{i\psiCM{\sigma}{j}}}.\]
Now~\eqref{eq:Field} evaluates to 
$H_\sigma = c_s\BP_\sigma\gamma_\sigma + c_n\BP_{\tau}\gamma_{\tau}$
and the bunch variables of each population evolve according to
{\allowdisplaybreaks
\begin{subequations}\label{eq:WS}
\begin{align}
\label{eq:WSrh}\dot\rho_\sigma &= \frac{1-\rho_\sigma^2}{2}\Re(H_\sigma e^{-i\Phi_\sigma}),\\
\label{eq:WSPh}\dot\Phi_\sigma &= \frac{1+\rho_\sigma^2}{2\rho_\sigma}\Im(H_\sigma e^{-i\Phi_\sigma}),\\
\label{eq:WSPs}\dot\Psi_\sigma &= \frac{1-\rho_\sigma^2}{2\rho_\sigma}\Im(H_\sigma e^{-i\Phi_\sigma}).
\end{align}
\end{subequations}}%
(The dynamics of individual oscillators~\eqref{eq:PhaseOsc} are determined by~\eqref{eq:WS} through \eqref{eq:PhaseOscField} and~\eqref{eq:Field}.)
Note that $\gamma_\sigma \to 1$ (and thus $z_\sigma\to Z_\sigma$) as $\maxdim\to\infty$ 
if the constants of motion are uniformly distributed on the circle, $\psiCM{\sigma}{k} = 2\pi k/\maxdim$, as shown in Ref.~\onlinecite{Pikovsky2008}; 
in this case $H_\sigma = c_s z_\sigma + c_n z_\tau$ and we recover~\eqref{eq:ComplxEqns} as~\eqref{eq:WSPs} decouples from~\eqref{eq:WSrh} and~\eqref{eq:WSPh}.

\begin{figure}
\includegraphics[width=\half\linewidth]{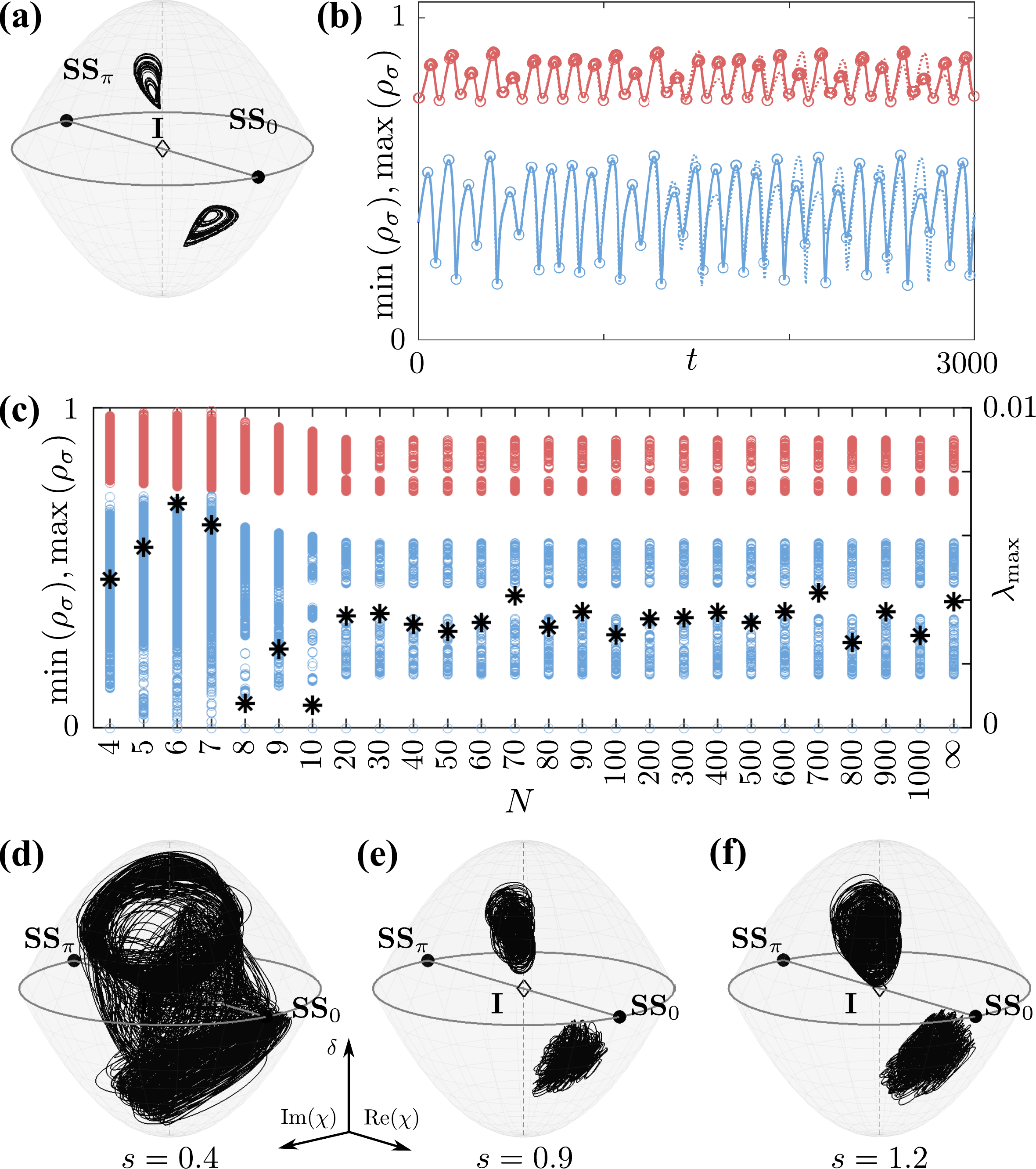}
\caption{
\label{fig:FiniteSize} 
Finite Kuramoto oscillator networks~\eqref{eq:WS} show robust chaos as the system size~$\maxdim$ and constants of motion, parametrized by~$s$, are varied; here $A = 0.7$, $\aln = 0.44$, $\als = 1.654$ (see Fig.~\ref{fig:Chaos}).
Panel~(a) shows the evolution of the bunch amplitudes $\rho_\sigma=\abs{z_\sigma}$ in~\eqref{eq:WS} for $\maxdim=20$, $s=1$. The projection $(\chi, \delta)= (z_1 \bar{z}_2, \abs{z_1}^2-\abs{z_2}^2)$ is analogous to that in Fig.~\ref{fig:Chaos}; they coincide for  $\maxdim\to\infty$ (for $s=1$).
Panel~(b) illustrates how the trajectory in Panel~(a) (solid lines) diverges from the dynamics of~$r_\sigma=\abs{\OP_\sigma}$ (dashed) in the continuum limit~\eqref{eq:cylgov}. Minima/maxima of the mean-field oscillations are highlighted (circles).
The observed chaotic dynamics is robust in~$s$ and~$\maxdim$:
Panel~(c) shows local minima/maxima in~$\abs{z_1}$ and~$\abs{z_2}$ (circles in Panel~(a)) 
and maximal Lyapunov exponent~$\lmax$ (asterisks) for varying network size~$\maxdim$ ($s=1$ fixed).
Panels~(d--f) shows projections of different chaotic attractors for $\maxdim=20$ as the constants of motion are varied through~$s$.
}
\end{figure}

Chaotic dynamics arise in networks of finitely many identical Kuramoto oscillators~\eqref{eq:PhaseOsc} for a wide range of system sizes. We fix phase lags $\als$, $\aln$ while varying~$\maxdim$ and take the constants of motion be uniformly distributed on the circle, $\psiCM{\sigma}{k} = 2\pi k/\maxdim$. The dynamics are now given by~\eqref{eq:WS}; effectively, these are the mean-field dynamics of the continuum limit~\eqref{eq:cylgov} modulated by finite-size fluctuations through~$\gamma_\sigma$ (which depend on~$\Psi_\sigma$ and vanish as $\maxdim\to\infty$). Fig.~\ref{fig:FiniteSize}(a,b) shows chaotic dynamics similar to those of the continuum limit (cf.~Fig.~\ref{fig:Chaos}) for $\maxdim=20$ oscillators per population. Numerical calculation of maximal Lyapunov exponent for varying system size, shown in Fig.~\ref{fig:FiniteSize}(c), indicates that there are not only chaotic dynamics for any network of $\maxdim\geq 20$ oscillators per population, but also for small networks.

The chaotic dynamics persist as the initial conditions are varied in the full system~\eqref{eq:PhaseOsc}. Keeping the constants of motion fixed will keep us on the same leaf of the foliation. But a generic perturbation of an initial conditions in the full system~\eqref{eq:PhaseOsc} will be on a different leaf of the foliation. To explore the dynamics for nearby leafs---and thus nearby initial conditions in~\eqref{eq:PhaseOsc}---we parametrize the constants of motion by $s\geq0$ by setting $\psiCM{\sigma}{k} = 2s\pi k/\maxdim$. Note that for $s=1$ we have a uniform distribution as above. Fig.~\ref{fig:FiniteSize}(d--f) shows the dynamics for varying parameter~$s$ for a network of $\maxdim=20$ oscillators per population. This suggests that even in small networks chaotic dynamics arise for many initial conditions.

\begin{figure}[b]
\includegraphics[width=\half\linewidth]{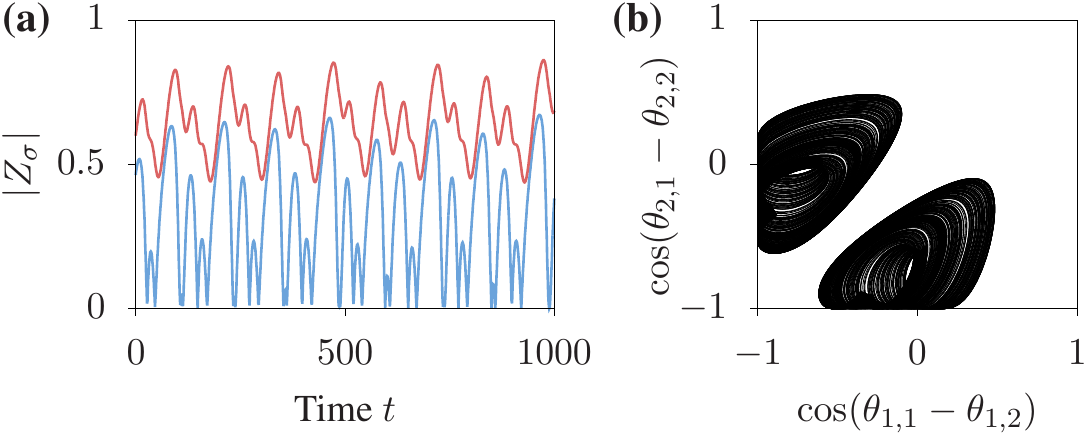}
\caption{\label{fig:SmallChaoticChimeras}Attracting chaos with $\lmax\approx 5.3275\cdot 10^{-3}$ arises in oscillator networks~\eqref{eq:PhaseOsc} of two populations of $\maxdim=2$ oscillators for parameters $A = 0.7$, $\als = 1.639$, and $\aln = 0.44$. Panel~(a) shows the evolution of the order parameters over time. Panel~(b) shows the phase evolution in a two-dimensional projection and a symmetric image.}
\end{figure}

There is further evidence that the mechanism that generates the chaotic dynamics is {universal} across system sizes, even where the mean-field reductions cease to apply. For nearby parameter values we find persistent chaos for two populations of~$\maxdim=2$ oscillators each; cf.~Fig.~\ref{fig:SmallChaoticChimeras}. This is the smallest network of two populations in which chaos can occur since the phase-space is effectively three-dimensional. These solutions are chaotic weak chimeras as defined in Refs.~\onlinecite{Ashwin2014a, Bick2015c, Bick2015d}: the asymptotic average frequencies $\Omega_{\sigma,k} := \lim_{T\to\infty}\frac{1}{T}\theta_{\sigma,k}(T)$ are the same with each population (due to symmetry) but distinct between populations. Hence our results also show that chaotic weak chimeras can occur even in the simplest system through symmetry breaking. A full analysis of this small system is beyond the scope of this manuscript and will be published elsewhere.

\section{Discussion.}%
Chaotic dynamics can---as conjectured by Ott and Antonsen~\cite{Ott2008}---indeed arise in two populations networks of coupled Kuramoto phase oscillators. Remarkably, these chaotic dynamics appear not only in the continuum limit and in large populations, but for roughly the same parameter values also in the smallest possible networks. While chaos has been observed in spatially extended (infinite-dimensional) mean-field equations~\cite{Wolfrum2016}, the setup of two populations is the smallest system possible in which chaos can arise in the mean field for Kuramoto oscillators. Moreover, the chaotic dynamics here are distinct from chaos in systems where interaction depends explicitly on the oscillators' phases (rather than the phase differences)~\cite{Marvel2009, Pazo2014} which have additional degrees of freedom. As in Ref.~\onlinecite{Bick2011}, chaos appears to relate to parameter values where the system has a time-reversal symmetry~\cite{Lamb1998}. Hence this raises the questions whether the symmetry induces suitable homoclinic or heteroclinic structures whose breaking yields attracting chaos across system sizes.

Our results show that in contrast to chaos induced by finite-size effects~\cite{Popovych2005}, there is chaos in the continuum limit for both identical ($\Delta=0$) and almost identical oscillators ($\Delta>0$) as given by the Ott--Antonsen reduction~\eqref{eq:ComplxEqns}. At the same time, we showed chaotic dynamics are also present in finite networks of identical oscillators whose dynamics are given by the Watanabe--Strogatz equations~\eqref{eq:WS}. However, neither of these approaches yields a suitable description of the finite-size networks of nonidentical oscillators; cf.~also Ref.~\onlinecite{Mirollo2012}. Is there chaos for finite networks of nonidentical oscillators? And if so, what are its properties, for example, the dimension of the attractor? Recently, perturbation theory has proved useful to describe the evolution of trajectories for near-integrable systems~\cite{Vlasov2016}, but new techniques are called for to describe the collective dynamics of nonidentical oscillator networks with respect to both the integrable case and the continuum limit.

In summary, oscillator networks~\eqref{eq:PhaseOsc} with simple sinusoidal interactions have surprisingly rich dynamics. For two populations of oscillators, higher-order effects such as amplitude variations or the influence of the fast oscillations, are not required to observe chaotic dynamics. Hence, we anticipate chaotic fluctuations to arise in small experimental oscillator setups~\cite{Bick2017, Martens2013}. Moreover, we expect much richer dynamics for three or more populations of phase oscillators~\cite{Martens2010c}. Such multi-population oscillator networks have been instructive to understand the dynamics of neural synchrony patterns~\cite*{Schmidt2014, Cabral2017}, where distributed phase lags are of particular importance due to the finite speed of signal propagation. Distributed phase lags give rise to chaotic dynamics and we therefore anticipate that our results further illuminate the dynamics of large-scale (neural) oscillator networks.

\section*{Acknowledgements}%
The authors would like to thank J~Engelbrecht, R~Mirollo, A~Politi, and M~Wolfrum for helpful discussions and F~Peter for careful reading of the manuscript.
CB would like to acknowledge the warm hospitality at DTU. Research conducted by EAM is partially supported by the Dynamical Systems Interdisciplinary Network, University of Copenhagen. CB has received partial funding from the People Programme (Marie Curie Actions) of the European Union's Seventh Framework Programme (FP7/2007--2013) under REA grant agreement no.~626111.

\bibliographystyle{apsrev4-1}
\def\urlprefix{}
\def\url#1{}

\bibliography{ref} 

\end{document}